From - Fri Oct 12 07:58:04 2001
Return-Path: <lemos@kelvin.ist.utl.pt>
Delivered-To: vcardoso@fisica.ist.utl.pt
Received: from kelvin.ist.utl.pt (kelvin.ist.utl.pt [193.136.161.154])
	by einstein.fisica.ist.utl.pt (Postfix) with ESMTP id 57FA54AAB4
	for <vcardoso@fisica.ist.utl.pt>; Thu, 11 Oct 2001 19:47:34 +0000 (GMT)
Received: (from lemos@localhost)
	by kelvin.ist.utl.pt (8.11.0/8.11.0) id f9BIhTm04557
	for vcardoso@fisica; Thu, 11 Oct 2001 19:43:29 +0100
Date: Thu, 11 Oct 2001 19:43:29 +0100
From: Sande Lemos <lemos@kelvin.ist.utl.pt>
Message-Id: <200110111843.f9BIhTm04557@kelvin.ist.utl.pt>
To: vcardoso@fisica.ist.utl.pt
X-Mozilla-Status: 8001
X-Mozilla-Status2: 00000000
X-UIDL: f53c9cc3e4fcf1f9

\documentstyle[12pt]{article}

\input{epsf.tex}

\begin{document}

  \begin{flushright} \begin{small}
  DF/IST-7.2001 \\gr-qc/0107098
  \end{small} \end{flushright}

\vskip 0.5cm

\begin{center}
{\bf QUASI-NORMAL MODES \\
OF TOROIDAL, CYLINDRICAL AND PLANAR\\
 BLACK HOLES IN ANTI-DE SITTER SPACETIMES:\\ SCALAR, ELECTROMAGNETIC AND 
GRAVITATIONAL PERTURBATIONS}
  \\
\vskip 1cm
Vitor Cardoso \\
\vskip 0.3cm
{\scriptsize  CENTRA, Departamento de F\'{\i}sica,
	      Instituto Superior T\'ecnico,}\\ 
{\scriptsize  Av. Rovisco Pais 1, 1096 Lisboa, Portugal,}\\
{\scriptsize E-mail: vcardoso@fisica.ist.utl.pt}\\
\vskip 0.6cm

Jos\'e P. S. Lemos \\
\vskip 0.3cm
{\scriptsize  CENTRA, Departamento de F\'{\i}sica,
	      Instituto Superior T\'ecnico,} \\
{\scriptsize  Av. Rovisco Pais 1, 1096 Lisboa, Portugal,}\\
{\scriptsize E-mail: lemos@kelvin.ist.utl.pt}
\end{center} 
 
\bigskip

\begin{abstract}
\noindent
We study the quasi-normal modes (QNM) of scalar, electromagnetic and
gravitational perturbations of black holes in general relativity whose
horizons have toroidal, cylindrical or planar topology in an
asymptotically anti-de Sitter (AdS) spacetime.  The associated
QNM frequencies describe the decay in time of the
corresponding test field in the vicinities of the  black hole.
In terms of the AdS/CFT conjecture, the inverse of the frequency is a
measure of the dynamical timescale of approach to thermal equilibrium
of the corresponding conformal field theory.
\strut  
\newline 
\end{abstract}

\noindent
\section{ Introduction}
\vskip 3mm

Black holes in anti-de Sitter (AdS) spacetimes in several dimensions
have been recently study. One of the reasons for this intense study is
the AdS/CFT conjecture which states that there is a correspondence
between string theory in AdS spacetime and a conformal field theory
(CFT) on the boundary of that space. For instance, M-theory on ${\rm
AdS}_4\times S^7$ is dual to a non-abelian superconformal field theory
in three dimensions, and type IIB superstring theory on ${\rm
AdS}_5\times S^5$ seems to be equivalent to a super Yang-Mills theory
in four dimensions \cite{maldacena,witten} (for a review see 
\cite{maldacenareview}).

All dimensions up to eleven are of interest in supersting theory, but
experiment singles out four dimensions as the most important. In
four-dimensional (4D) general relativity, an effective gravity theory 
in an appropriate  
string theory limit, the Kerr-Newman family of four-dimensional 
black holes can be extended to include a negative cosmological
constant \cite{bcarter}.  The horizon in this family has the topology
of a sphere. There are, however, other families of black holes in general
relativity with a negative cosmological constant with horizons having
topology different from spherical. Here we want to focus on the family
of black holes whose horizon has toroidal, cylindrical or planar
topology \cite{lemos1,lemos2,lemoszanchin96} (see \cite{lemosreview}
for a review).  We are going to perturb these black holes with 
scalar, electromagnetic and gravitational fields.

Perturbations of known solutions are very important to perform in
order to study their intrinsic properties, such as the natural
frequencies of the perturbations, and to test for the stability of the
solutions themselves. For gravitational objects, such as a black hole,
the vibrational pattern set by the perturbation obliges the system to
emit gravitational waves. Thus, for black holes, the study of
perturbations is closely linked to the gravitational wave emission.
Due to the dissipative character of the emission of the gravitational
waves the vibrational modes do not for a normal set, indeed in the
spectrum each frequency is complex whose imaginary part gives the
damping timescale. These modes are called quasi-normal modes (QNMs).
The QNMs of a black hole appear naturally when one deals with the
evolution of some field in the black hole spacetime, and serve as a
probe to the dynamics outside its event horizon.

Much work has been done for black holes in asymptotically flat
spacetimes (see \cite{kokkotas99} for a review), the main interest
being related to the gravitational waves emitted when these
astrophysical objects form.  In turn, the recent AdS/CFT
correspondence conjecture has attracted much attention to the
investigation of QNMs in anti-de Sitter spacetimes. According to it,
the black hole corresponds to a thermal state in the conformal field
theory, and the decay of the test field in the black hole spacetime,
correlates to the decay of the perturbed state in the CFT.  The
dynamical timescale for the return to thermal equilibrium can be
computed using the black hole characteristics plus the AdS/CFT
correspondence. Since one can always compactify extra unused
dimensions, the study of black holes in any permitted dimension (from
2 to 11) is useful. In 3D QNMs were studied in
\cite{chanmann1,Govinda,birmingham,cardoso1} (see also \cite{Satoh})
 for scalar perturbations and in \cite{cardoso1} for Maxwell and Weyl
perturbations. In 4D QNMs were studied in \cite{hubeny,horowitz,Wang3}
for Schwarzschild-AdS black holes with minimally coupled scalar field
perturbations and box type boundary conditions and in \cite{chanmann1}
with conformally coupled scalar field and asymptotically flat boundary
conditions, in \cite{Wang1,Wang2,kim} for Reissner-Nordstr\"om black
holes with scalar perturbations, in \cite{cardoso2} for
Schwarzschild-AdS black holes with Maxwell and gravitational
perturbations, and in \cite{chanmann2} for topological black holes
with scalar perturbations. For higher dimensions, such as 5 and 7D see
\cite{Govinda,hubeny}, and for a recent work on super-radiance in the
Kerr-Newman-anti-de Sitter geometry see \cite{win}.

In this paper we shall study scalar, electromagnetic and gravitational
perturbations of the toroidal, cylindrical or planar black holes in an
AdS spacetime found in \cite{lemos2}.  The motivation to perturb with
a scalar field can be seen as follows.  If one has, e.g.,
11-dimensional M-theory, compactified into a $({\rm toroidal\;
BH})_{4} \times ({\rm compact\; space})$ the scalar field used to
perturb the black hole, can be seen as a type IIB dilaton which
couples to a CFT field operator $\cal{O}$. Now, the black hole in the
bulk corresponds to a thermal sate in the boundary CFT, and thus the
bulk scalar perturbation corresponds to a thermal perturbation with
nonzero $<\cal O>$ in the CFT.  Similar arguments hold for the
electromagnetic perturbations since they can be seen as perturbations
for some generic gauge field in the low energy limit of 11-dimensional
M-theory.  On the other hand, gravitational perturbations are always
of importance since they belong to the essence of the spacetime
itself.

We will find that the QNM frequencies for scalar perturbations scale
with the horizon radius, at least for large black holes.  In the case
of electromagnetic perturbations of large black holes, the
characteristic QNM frequencies have only an imaginary part, and scale
with the horizon radius.  As for gravitational perturbations, there
are two important features. First, contrary to the asymptotically flat
spacetime case, odd and even perturbations no longer have the same
spectra, although in certain limits one can still prove that the
frequencies are almost the same.  The second  result is
that, for odd perturbations, there is a mode with a totally different
behavior from that found in the scalar and electromagnetic case: in
this mode the frequency scales with $\frac{1}{r_+}$, just as in
asymptotically flat Schwarzschild spacetime. 
These features were also found in our study of Schwarzschild-AdS 
black holes \cite{cardoso2}. One could have predicted 
that these features would also appear here, at leat for large 
balck holes, because in the large horizon limit 
the spherical-AdS black holes have the geometry of the 
cylindrical (planar or toroidal) ones.

\vskip 0.5cm

\noindent
\section{Scalar, electromagnetic and gravitational perturbations in a 
toroidal, cylindrical or planar black hole in an  AdS background}

\vskip 3mm
Throughout this paper, we shall deal with the evolution of some 
perturbation in a spacetime geometry in general relativity 
with a background metric given by 
\cite{lemoszanchin96}: 
\begin{equation}
ds^{2}= f(r)\,dt^{2}- f(r)^{-1}dr^{2}-r^2\,dz^{2}
-r^{2}d\phi^{2}\,
\label{lineelement}
\end{equation}
where 
\begin{equation}
f(r)=\frac{
r^2}{R^2}-\frac{4MR}{r}\, ,
\label{f(r)}
\end{equation}
$M$ is the ADM mass of the black hole, 
and $R$ is the AdS lengthscale $R^2=-\frac{3}{\Lambda}$, $\Lambda$ being
the cosmological constant. There is a horizon at $r_+=(4M)^{1/3}R$.
The range of the coordinates $z$ and $\phi$ dictates the topology of
the black hole spacetime.  For a black hole with toroidal topology, a
toroidal black hole, the coordinate $z$ is compactified such that
$z/R$ ranges from $0$ to $2\pi$, and $\phi$ ranges from $0$ to $2\pi$
as well.  For the cylindrical black hole, or black string, the
coordinate $z$ has range $-\infty<z<\infty$, and $0\leq \phi
<2\pi$. For the planar black hole, or black membrane, the coordinate
$\phi$ is further decompactified $-\infty<R\,\phi<\infty$
\cite{lemoszanchin96}.  We will work with the cylindrical topology but
the results are not altered for the other two topologies. 

According to the AdS/CFT correspondence solution 
(\ref{lineelement}) ($\times S^7$) is dual to a superconformal field theory 
in three dimensions with ${\cal N}=8$. 
These toroidal black holes with Ricci flat 
horizon we are considering in (\ref{lineelement}) can be seen as 
the large horizon radius limit, $r_+/R>>1$, of the spherical AdS black holes.
The large black holes are the ones that matter most to the AdS/CFT 
correspondence 
\cite{witten}.
Perturbations of spherical black holes were studied in \cite{cardoso2} (see 
also references in the Introduction). Therefore the results we obtain here 
should be similar to the results we have obtained in \cite{cardoso2} for 
large black holes. This will be confirmed below. For small black holes 
we will show that the spherical and toroidal yield different results.

\subsection{Scalar field perturbations}


For scalar perturbations, we are interested in solutions to 
the minimally coupled scalar wave equation 
\begin{equation}
{\Phi^{,\,\mu}}_{;\,\mu}=0 \,, 
\label{minimalscalareq1}
\end{equation}
where, a comma stands for ordinary derivative and a semi-colon stands
for covariant derivative.
We make the following ansatz for the field $\Phi$ 
\begin{equation}
\Phi=\frac{1}{r}P(r)e^{-i\omega t}e^{ikz}e^{il\phi}\,,
\label{ansatzforscalar}
\end{equation}
where $\omega$, $k$, and $l$, are the frequency, the wavenumber and
the angular quantum numbers of the perturbation. If one is dealing 
with the toroidal  topology then $k$ should be changed into  
an angular quantum number $\bar l$,   
$ e^{ikz} \rightarrow e^{i{\bar l}\frac{z}{R}}$. For the planar 
topology $e^{il\phi}\rightarrow e^{i{\bar k}R\,\phi}$, where 
$\bar k$ is now a continuous wave number. 

It is useful to use the tortoise 
coordinate  $r_*$
defined by the equation $dr_*=dr/\left({r^2}/{R^2}-{4MR}/{r}\right)$. 
With the ansatz (\ref{ansatzforscalar}) and the tortoise coordinate $r_*$, 
equation 
(\ref{minimalscalareq1}) is given by, 
\begin{equation}
\frac{d^2 P(r)}{d {r_*}^2} + \left\lbrack\omega^2 - 
V_{\rm scalar}(r)\right\rbrack
P(r)=0\,,
\label{minimalscalareq2}
\end{equation}
where,
\begin{equation}
V_{\rm scalar}(r)=f\left(\frac{l^2}{r^2}+
\frac{R^2k^2}{r^2}+\frac{f'}{r}\right)\,,
\label{potentialscalar}
\end{equation}
with $r=r(r_*)$ given implicitly and $f'\equiv df/dr$.
The rescaling to the radial coordinate 
$\hat{r}=\frac{r}{R}$ and to the frequency $\hat{\omega}=\omega R$ is 
equivalent 
to take $R=1$ in (\ref{minimalscalareq2}) and (\ref{potentialscalar}), i.e., 
through this rescaling one measures the frequency and other quantities in 
terms of the AdS lengthscale $R$.

\subsection{Maxwell field perturbations}
We consider the evolution of a Maxwell field in a
cylindrical-AdS black hole spacetime with a metric
given by (\ref{lineelement}).
The evolution
is governed by Maxwell's equations:
\begin{equation}
{F^{\mu\nu}}_{;\nu}=0\,,\quad F_{\mu\nu}=A_{\nu,\mu}-A_{\mu,\nu}\,.
\label{maxwell} 
\end{equation}
One can again separate variables by the ansatz:
\begin{eqnarray}
A_{\mu}(t,r,\phi,z)=
\left[ \begin{array}{c}g^{k\,l}(r)\\h^{k\,l}(r)
\\k^{k\,l}(r) \\ j^{k\,l}(r)\end{array}\right]
e^{-i\omega t}
e^{ikz}
e^{il\phi} \,, 
\label{empotentialdecomposition}
\end{eqnarray}
When we put this expansion into Maxwell's
equations (\ref{maxwell}) we get the $2^{\rm nd}$ order differential
equation for the perturbation:
\begin{equation}
\frac{\partial^{2} \Psi(r)}{\partial r_*^{2}} +\left\lbrack\omega^2
-V_{\rm maxwell}(r)\right\rbrack\Psi(r)=0 \,,
\label{wavemaxwell}
\end{equation}
where the wavefunction $\Psi$ is a  combination of the functions
$g^{lm}$, $h^{lm}$, $k^{lm}$ and $j^{lm}$ as appearing in
(\ref{empotentialdecomposition}), and $\Psi=-i\omega h^{lm}-
\frac{dg^{lm}}{dr}$ (see \cite{Ruffini} for further details).
The potential $V_{\rm maxwell}(r)$ appearing in equation (\ref{wavemaxwell}) 
is given by
\begin{equation}
V_{\rm maxwell}(r)=f(r)\left(\frac{l^2}{r^2}
+\frac{R^2k^2}{r^2}\right) \,.
\label{potentialmaxwell}
\end{equation}
Again we can
take $R=1$ and measure everything in terms of $R$.
\subsection{ Gravitational perturbations}
In our analysis of gravitational perturbations, we shall adopt a
 procedure analogous to that of Chandrasekhar \cite{Chandra2}, generalizing
the calculation by the introduction of the cosmological constant 
$\Lambda=-\frac{3}{R^2}$.
The perturbed metric will be taken to be
\begin{equation}
ds^{2}= e^{2\nu}dt^{2}- e^{2\psi}(d\phi-wdt-q_2dr-q_3dz)^2-
e^{2\mu_2}dr^2-e^{2\mu_3}dz^2\,.
\label{lineelementchandra}
\end{equation}
where the unperturbed quantities are $e^{2\nu}=\frac{ r^{2}}{R^2}-
\frac{4MR}{r}$,
$e^{2\psi}=r^{2}$, $e^{2\mu_2}=(\frac{
r^2}{R^2}-\frac{4MR}{r})^{-1}$, $e^{2\mu_3}=\frac{r^2}{R^2}\,$ and all the 
other unperturbed quantities are zero.
By observing the effect of performing $\phi\rightarrow-\phi$, and
maintaining the nomenclature of the spherical symmetric case
\cite{Chandra2}, it can be seen that the perturbations fall into two
distinct classes: the odd perturbations (also called axial in the
spherical case) which are the quantities $w$, $q_2$, and $q_3$, and
the even perturbations (also called polar in the spherical case) which
are small increments $\delta\nu$, $\delta\mu_2$, $\delta\mu_3$ and
$\delta\psi$ of the functions $\nu$, $\mu_2$, $\mu_3$ and $\psi$,
respectively.  Note that since $z$ is also an ignorable coordinate,
one could, in principle, interchange $\phi$ with $z$ in the above argument. 

In what follows, we shall limit ourselves to  axially symmetric
perturbations, i.e., to the case in which the quantities listed above
do not depend on $\phi$. In this case odd and even perturbations
decouple, and it is possible to simplify them considerably.

\subsubsection{Odd perturbations}
We will deal with odd perturbations first. As we have stated, these
are characterized by the non vanishing of $w\,$, $q_2\,$ and
$q_3$. The equations governing these quantities are the following Einstein's
equations with a cosmological constant
\begin{equation}
G_{12}+\frac{3}{R^2}g_{12}=0\,,
\label{g01}
\end{equation}
\begin{equation}
G_{23}+\frac{3}{R^2}g_{23}=0\,.
\label{g12}
\end{equation}
The reduction of these two equations to a one dimensional second order
differential equation is well known (see \cite{Chandra2}), and we only
state the results. By defining
\begin{equation}
Z^-(r)=r\left(\frac{ r^{2}}{R^2}-\frac{4MR}{r}\right)
\left(\frac{dq_2}{dz}-\frac{dq_3}{dr}\right)\,,
\label{Z-}
\end{equation}
One can easily check that $Z^-(r)$ satisfies
\begin{equation}
\frac{\partial^{2} Z^-(r)}{\partial r_*^{2}} +\left\lbrack\omega^2
-V_{\rm odd}(r)\right\rbrack Z^-(r)=0 \,,
\label{waveaxial}
\end{equation}
where $V_{\rm odd}(r)$ appearing in equation (\ref{waveaxial}) is given by
\begin{equation}
V_{\rm odd}(r)=f(r)\left(\frac{R^2k^2}{r^2}-\frac{12MR}{r^3}\right) \,.
\label{potentialaxial}
\end{equation}

\subsubsection{Even perturbations}
Even perturbations are characterized by non-vanishing increments 
in the metric functions $\nu\,$, $\mu_2\,$, $\mu_3\,$ and $\psi\,$.
The equations which they obey are obtained by linearizing 
$G_{01}+\frac{3}{R^2}g_{01}$,
$G_{03}+\frac{3}{R^2}g_{03}$, 
$G_{13}+\frac{3}{R^2}g_{13}$, 
$G_{11}+\frac{3}{R^2}g_{11}$ and
$G_{22}+\frac{3}{R^2}g_{22}$ about their unperturbed values.
Making the ansatz
\begin{eqnarray}
\delta \nu=N(r)e^{ikz}\,,\\
\delta \mu_2=L(r)e^{ikz}\,,\\
\delta \mu_3=T(r)e^{ikz}\,, \\ 
\delta \psi=V(r)e^{ikz}  \,, 
\label{ansatzpolar}
\end{eqnarray}
we have from $\delta(G_{03}+\frac{3}{R^2}g_{03})=0$ that
\begin{equation}
V(r)=-L(r)\,.
\label{relpol1}
\end{equation}
Inserting this relation and using (\ref{ansatzpolar}) in 
$\delta(G_{01}+\frac{3}{R^2}g_{01})=0\,$,
we have  
\begin{equation}
\left(\frac{3}{r}-\frac{f'}{2f}\right)L(r)
+\left(\frac{f'}{2f}-\frac{1}{r}\right)T(r)+L(r)_{,r}-
T(r)_{,r}=0\,.
\label{A}
\end{equation}
 From $\delta(G_{13}+\frac{3}{R^2}g_{13})=0\,$ we have
\begin{equation}
\left(\frac{1}{r}+\frac{f'}{2f}\right)L(r)+
\left(\frac{1}{r}-\frac{f'}{2f}\right)N(r)-N(r)_{,r}+
L(r)_{,r}=0\,,
\label{C}
\end{equation}
and from $\delta(G_{11}+\frac{3}{R^2}g_{11})=0\,$ we obtain
\begin{eqnarray}
&-\frac{k^2R^2}{fr^2}N(r)+
\left(\frac{k^2R^2}{fr^2}-\frac{\omega^2}{f^2}-\frac{6}{f}\right)L(r)+
\frac{\omega^2}{f^2}T(r)+
&\nonumber\\&
\left(\frac{1}{r}+\frac{1}{Rr}\right)N(r)_{,r}+
\left(-\frac{f'}{2f}-\frac{1}{Rr}\right)L(r)_{,r}
+ \left(\frac{f'}{2f}+\frac{1}{r}\right)T(r)_{,r}=0\,.
\label{D}
\end{eqnarray}
Multiplying equation (\ref{C}) by $\frac{2}{r}$ and adding (\ref{D}) we 
can obtain $N(r)$
and $N(r)_{,r}$ in terms of $L(r)$, $L(r)_{,r}$, $T(r)$ and $T(r)_{,r}$. 
Using (\ref{A}) and (\ref{C}) we can express $L(r)\,$, $N(r)\,$, and up to
their second derivatives in terms of $T(r)$, $T(r)_{,r}$ and $T(r)_{,rr}$.
Finally, we can look for a function
\begin{equation}
Z^+=a(r)T(r)+b(r)L(r)\,.
\label{defZ}
\end{equation}
which satisfies the second order differential equation
\begin{equation}
\frac{\partial^{2} Z^+(r)}{\partial r_*^{2}} +\left\lbrack\omega^2
-V_{\rm even}(r)\right\rbrack Z^+(r)=0 \,,
\label{wavepolar}
\end{equation}
Substituting (\ref{defZ}) into (\ref{wavepolar}) and expressing $L(r)$ and
its derivatives in terms of $T(r)$ and its derivatives, we obtain an equation
in $T(r)$, $T(r)_{,r}$ and $T(r)_{,rr}$ whose coefficients must vanish 
identically.
If we now demand that $a(r)$ and $b(r)$ do not depend on the frequency 
$\omega$ we find 
\begin{eqnarray}
a(r)=\frac{r}{12Mr+k^2r^2}\,,\\
b(r)=\frac{6M+k^2r}{72M^2+6k^2Mr}\,,
\label{relation}
\end{eqnarray}
and the potential $V_{\rm even}(r)$ in (\ref{wavepolar}) is
\begin{equation}
V_{\rm even}(r)=f(r)\left\lbrack\frac{576M^3+12k^4Mr^2+k^6r^3+
144M^2r(k^2+2r^2)}{r^3(12M+k^2r)^2}\right\rbrack \,.
\label{potentialpolar}
\end{equation}

As a final remark concerning the wave equations obeyed by odd and even
gravitational perturbations, we note that it can easily be checked that the 
two 
potentials can be expressed in the form
\begin{eqnarray}
V_{\stackrel{\rm odd}{\rm even}}= 
W^2 \pm \frac{dW}{dr_*} +\beta \,, \\
W=\frac{96M^2(k^2+3r^2)}{2k^2r^2(12M+k^2r)}+j\,,
\label{W}
\end{eqnarray}
where $j=-\frac{k^6+288M^2}{24k^2M}$, and
$\beta=-\frac{k^8}{576M^2}$.  It is worth of notice that the two
potentials can be written in such a simple form (potentials related in
this manner are sometimes called superpartner potentials
\cite{Cooper}), a fact which seems to have been discovered by
Chandrasekhar \cite{Chandra2}.

\vskip 0.5cm

\noindent
\section{Quasi-normal modes and some of its properties}

\vskip 3mm


\subsection{Boundary conditions}
To solve (\ref{minimalscalareq2}), (\ref{wavemaxwell}),
(\ref{waveaxial}) and (\ref{wavepolar}) one must specify boundary
conditions, a non-trivial task in AdS spacetimes.  Consider first the
case of a Schwarzschild black hole in an asymptotically flat spacetime
(see \cite{kokkotas99}). Since the potential now vanishes at both
infinity and the horizon, two independent solutions near these points
are $ \Psi_1 \sim e^{-i\omega r_*}$ and $\Psi_2 \sim e^{i\omega r_*}$,
where the $r_*$ coordinate now ranges from $-\infty $ to $\infty$.
QNMs are defined by the condition that at the horizon
there are only ingoing waves, $\Psi_{\rm hor}\sim e^{-i\omega r_*} $.
Furthermore, one wishes to have nothing coming in from infinity (where
the potential now vanishes), so one wants a purely outgoing wave at
infinity, $\Psi_{\rm infinity}\sim e^{i\omega r_*} $. Clearly, only a
discrete set of frequencies $\omega$ meet these requirements.

Consider now our asymptotically AdS spacetime.  The first boundary
condition stands as it is, so we want that near the horizon
$\Psi_{\rm hor}\sim e^{-i\omega r_*} $. However $r_*$ has a finite range,
so the second boundary condition needs to be changed. There have been
a number of papers on which boundary conditions to impose at infinity
in AdS spacetimes (\cite{Avis}-\cite{Burgess}).  We
shall require energy conservation and adopt reflective boundary
conditions at infinity \cite{Avis}
which means that the wavefunction is zero at infinity
(see however \cite{Dasgupta}).

\subsection{Numerical calculation of the QNM frequencies}
To find the frequencies $\omega$ that satisfy the previously stated
boundary conditions we first change wavefunction to $\phi=e^{i\omega r_*} Z$ 
(where, $Z=P\,,\Psi\,, Z^+\, ,Z^-$).
The wave equation then transforms into 
\begin{equation}
f(r)\frac{\partial^{2} \phi}{\partial r^{2}}+
\left(f'-2i\omega\right)\frac{\partial \phi}{\partial r} -\frac{V}{f}\phi=0\,.
\label{waveeq}
\end{equation}
We now note that (\ref{waveeq}) has only regular
singularities in the range of interest. It has therefore, by Fuchs
theorem, a polynomial solution. To deal with the point at infinity, we
first change the independent variable to $x=\frac{1}{r}$. Now we can use
Fr\"{o}benius method by looking for an indicial equation (for further
details see \cite{hubeny}), and force it to obey the boundary condition
at the horizon ($x=\frac{1}{r_+}=h$).  We get \begin{eqnarray} Z(x)=
\sum_{j=0}^{\infty} a_{j(\omega)} (x-h)^j \,,
\label{frobenius} 
\end{eqnarray}
where $a_{j(\omega)}$ is a function of the frequency.  If we put
(\ref{frobenius}) into (\ref{waveeq}) and use the boundary condition
$Z=0$ at infinity ($x=0$) we get:
\begin{equation}
\sum_{j=0}^{\infty} a_{j(\omega)}(-h)^j=0
\label{numerico} 
\end{equation}
Our problem is reduced to that of finding a 
numerical solution of the polynomial equation (\ref{numerico}).
The numerical roots for $\omega$ of equation (\ref{numerico}) can be
evaluated, resorting to numerical computation. Obviously, one cannot
determine the full sum in expression ($\ref{numerico}$), so we have to
determine a partial sum from $0$ to $N$, say and find the roots $\omega$
of the resulting polynomial expression.  We then move onto the next
term $N+1$ and determine the roots.  If the method is reliable, the
roots should converge. We stop our search once we have a 3 decimal digit
precision. One can label the roots $\omega$ with the principal quantum 
number $n$, $\omega_n$. We have only looked for the roots with lowest 
imaginary part, e.g. $n=1,2$. The $n=1$ solution is
called  the lowest QNM. By default 
we write the frequencies for the lowest $n=1$ and first excited 
$n=2$ QNMs as $\omega$ simply, instead of $\omega_1,\, \omega_2$.

As we will see there are frequencies with a vanishing real
part, which makes it possible to use an approximation, due to Liu, to
these highly damped modes \cite{Liu1,Liu2}. Although the method was
originally developed for the asymptotically flat space, it is quite
straightforward to apply it to our case.  There is therefore a way to
test our results. Unfortunately, this method relies heavily
on having not only a pure imaginary frequency
but also a frequency with a large imaginary part, so as we shall see
it will only work for electromagnetic perturbations.
  We have computed the lowest frequencies for some
values of the horizon radius $r_+$, and $l$.  
The frequency is written as $\omega = \omega_r +
i\omega_i$, where $\omega_r$ is the real part of the frequency and
$\omega_i$ is its imaginary part. We present the results in tables 1-4.

\vskip 0.5cm
\subsubsection{Scalar:}
\vskip 0.5cm
\begin{center}
\begin{tabular}{|l|l|l|l|l|}  \hline 
\multicolumn{1}{|c|}{} &
\multicolumn{2}{c|}{ Numerical} \\ \hline
$r_+$    &  $-\omega_i$ &  $\omega_r$ \\ \hline
0.1  & 0.266 & 0.185   \\ \hline
 1   &  2.664  &  1.849   \\ \hline
5  & 13.319 & 9.247  \\ \hline
10  & 26.638  & 18.494    \\ \hline
50  &133.192 & 92.471   \\ \hline
100 &  266.373 &184.942  \\ \hline
\end{tabular}
\end{center}
\vskip 1mm
{\noindent Table 1. Lowest QNM of scalar perturbations for $l=0\,$ and $k=0$.}
\vskip 1cm

In table 1 we list the numerical values of the lowest ($n=1$) QNM
frequencies for the $l=0$ scalar field QNM and for selected values of
$r_+$. As discussed in Horowitz and Hubeny (HH) \cite{hubeny} the
frequency should be a function of the scales of the problem, $R$ and
$r_+$. However, they showed and argued that due to additional
symmetries in the scalar field case and for large Schwarzschild-AdS
black holes the frequency scales as $\omega\propto T$, with the
temperature of the large black hole given by $T\propto r_+/R^2$, i.e.,
$\propto r_+$ in our units. This behavior is a totally different
behavior from that of asymptotically flat space, in which the
frequency scales with $\frac{1}{r_+}$.  Now, for cylindrical (planar
or toroidal) black holes and scalar fields this symmetry is present
for any horizon radius, so $\omega\propto r_+$ always.  For these
black holes the temperature is also proportional to $r_+$, no matter
how small the black hole is \cite{lemos1,peca}. Thus, the scalar field
QNM frequencies are proportional to $T$, as can be directly seen from
table 1, $\omega\propto r_+\propto T$.  The imaginary part of the
frequency determines how damped the mode is, and according to the
AdS/CFT conjecture is a measure of the characteristic time
$\tau=\frac{1}{\omega_i}$ of approach to thermal equilibrium in the
dual CFT (moreover, the frequencies do not seem to depend on the
angular quantum number $l$, we have performed calculations for higher
values of $l$).  In the dual CFT the approach to thermal equilibrium
is therefore faster for higher temperatures, i.e., larger black holes.
This scaling for all horizon radii with temperature only happens in
the scalar field case.  For the electromagnetic and some of the
gravitational perturbations the frequency scales with the temperature
only in the large black hole regime, as we will show.

In table 2 we list the numerical values of the lowest ($n=1$)
QNM frequencies for $l=1$ and for selected values of
$r_+$.  For frequencies with no real part, we list the values
obtained in the ``highly damped approximation'' \cite{Liu1,Liu2}. 
We can note from table 2 that $\omega$ is proportional to $r_+$ and thus to 
the temperature for large black holes, $r_+ \buildrel>\over\sim  
5$, say.

\vskip 0.5cm
\subsubsection{Electromagnetic:}
\vskip 0.5cm
\begin{center}
\begin{tabular}{|l|l|l|l|l|}  \hline 
\multicolumn{1}{|c|}{} &
\multicolumn{2}{c|}{ Numerical} &
\multicolumn{2}{c|}{ Highly Damped} \\ \hline
$r_+$    &  $-\omega_i$ &  $\omega_r$ &  $-\omega_i$  & $\omega_r$ \\ \hline
0.1  & 0.104 &1.033  & $-$ & $-$  \\ \hline
 1   &  1.709  & 1.336  & $-$ & $-$   \\ \hline
5  &7.982  & $\sim0$ & 7.500 & $\sim0$  \\ \hline
10  &15.220   & $\sim0$  & 15.000 & $\sim0$  \\ \hline
50  &75.043 & $\sim0$ & 75.000 & $\sim0$  \\ \hline
100 & 150.021  & $\sim0$ & 150.000 & $\sim0$  \\ \hline
\end{tabular}
\end{center}
\vskip 1mm
{\noindent Table 2. Lowest QNM of electromagnetic perturbations for $l=1\,$ 
and $k=0$.}
\vskip 1.2cm

In the large black hole regime, both scalar and electromagnetic QNM
frequencies for this geometry are very similar to those of the
Schwarzschild-anti-de Sitter black hole \cite{hubeny,cardoso2}. This
is a consequence of the fact that in this regime, the wave equation
for the fields become identical in both geometries.  Indeed we can
compare table 1 for the scalar field with table 1 of 
HH  \cite{hubeny} for the same field. We see that for $r_+=1$
one has for the toroidal black hole $\omega=1.849-2.664i$, whereas HH
find $\omega=2.798-2.671 i$. Thus they differ, $r_+=1$ is not a large
black hole. For $r_+>>1$ one expects to find very similar QNM
frequencies.  For instance, for $r_+=100$ we obtain
$\omega=184.942-266.373i$ for the toroidal black hole, while HH obtain
$\omega=184.953-266.385 i$ for the Schwarzschild-anti de Sitter black
hole.  
The same thing happens for electromagnetic perturbations.  We
can compare table 2 for the electromagnetic field with table I of
Cardoso and Lemos \cite{cardoso2}. For large black holes one can see
that the frequencies for toroidal black holes are very similar to the
frequencies of the Schwarzschild-anti-de Sitter black hole. For
instance, for $r_+=100$ we find $\omega=-150.021i$ for the toroidal
black hole, while in \cite{cardoso2} we found $\omega=-150.048 i$ for
the Schwarzschild-anti-de Sitter black hole.  We can also see that the
QNM modes in the electromagnetic case are in excellent agreement with
the analytical approximation for strongly damped modes.

\subsubsection{Gravitational:}
The numerical calculation of the QNM frequencies for gravitational
perturbations proceeds as outlined previously (the associated
differential equation has only regular singularities, so it is
possible to use an expansion such as (\ref{frobenius})). In tables 3
and 4 we show the two lowest lying ($n=1\,,2$) QNM frequencies for
$l=2$ and $l=3$ gravitational perturbations.

We first note that there is clearly a distinction between odd and even
perturbations: they no longer have the same spectra, contrary to the
asymptotically flat space case (see \cite{Chandras}).  This problem
was studied in some detail by Cardoso and Lemos \cite{cardoso2} who
showed that it is connected with the behavior of $W$ (see equation
(\ref{W})) at infinity.
\vskip 1.5cm
\noindent 
{\bf odd (axial) modes:}
\vskip 1cm
\begin{center}
\begin{tabular}{|l|l|l|l|l|}  \hline 
\multicolumn{1}{|c|}{} &
\multicolumn{2}{c|}{ lowest QNM} &
\multicolumn{2}{c|}{ second lowest QNM} \\ \hline
$r_+$    &  $-\omega_i$ &  $\omega_r$ &  $-\omega_i$  & $\omega_r$ \\ \hline
1  & 2.646  & $\sim0$  &2.047  & 2.216   \\ \hline
5  & 0.2703 & $\sim0$ &13.288  &9.355   \\ \hline
10   &  0.13378 & $\sim0$  & 26.623 & 18.549  \\ \hline
50   & 0.02667 & $\sim0$ & 133.189 & 92.482 \\ \hline
100  & 0.0134  & $\sim0$ & 266.384 & 184.948  \\ \hline
\end{tabular}
\end{center}
\vskip 1mm
{\noindent Table 3. Lowest and second lowest 
QNMs of gravitational odd perturbations for 
$k=2$}.
\vskip 2.5cm
\noindent 
{\bf even (polar) modes:}
\vskip 0.5cm
\begin{center}
\begin{tabular}{|l|l|l|l|l|}  \hline 
\multicolumn{1}{|c|}{} &
\multicolumn{2}{c|}{ lowest QNM, $k=2$} \\ \hline
$r_+$    &  $-\omega_i$ &  $\omega_r$ \\ \hline
1   & 1.552 & 2.305   \\ \hline
5  &  12.633 & 9.624     \\ \hline
10  & 26.296  & 18.696     \\ \hline
50  & 133.124 & 92.512       \\ \hline
100  & 266.351  & 184.963     \\ \hline
\end{tabular}
\end{center}
\vskip 1mm
{\noindent Table 4. Lowest QNM of gravitational even perturbations for 
$k=2$.}
\vskip 1cm

For odd gravitational QNMs the lowest one scales with
$\frac{1}{r_+}\propto \frac{1}{T}$. This is odd, but one can see that
it is a reflection of the different behavior of the potential $V_{\rm
odd}$ for odd perturbations.  For the second lowest odd gravitational
QNM table 3 shows that for large black holes it scales with $T$.  For
the lowest even gravitational QNM table 4 shows that it also scales
with $T$ for large black holes.  Note further that the scalar, odd
second lowest and even gravitational QNMs are very similar in the
large black hole regime.  Indeed, tables 1, 3 and 4 show a remarkable
resemblance even though the potentials are so different.  Finally,
let us compare tables 3 and 4 with tables III-V of \cite{cardoso2}.  We
see that for large black holes the frequencies of toroidal black holes
are again very similar to those of the Schwarzschild-anti-de Sitter
black hole.  For instance, for odd perturbations and $r_+=100$ we find
from table 3 $\omega=-0.0134i$ for the toroidal black hole, while in
\cite{cardoso2} we found (table III) $\omega=-0.0132i$ for the
Schwarzschild-anti-de Sitter black hole. 

\vskip 0.5cm

\noindent
\section{Conclusions}
\vskip 3mm

We have computed the scalar, electromagnetic and gravitational QNM
frequencies of the toroidal, cylindrical or planar black hole in four
dimensions. These modes dictate the late time behaviour of a minimally
coupled scalar, electromagnetic field and of small gravitational
perturbations, respectively.  The main conclusion to be drawn from
this work is that these black holes are stable with respect to small
perturbations.  In fact, as one can see, the frequencies all have a
negative imaginary part, which means that these perturbations will
decay exponentially with time.  For odd gravitational perturbations in
the large black hole regime, the imaginary part of the frequency goes
to zero scaling with $\frac{1}{r_+}$, just as in asymptotically flat
space and in the odd gravitational perturbations of Schwarzschild-Ads
black hole.  In terms of the AdS/CFT correspondence, this implies that
the greater the mass, the more time it takes to approach equilibrium,
a somewhat puzzling result.  Apart from this interesting result, the
frequencies all scale with the horizon radius, at least in the large
black hole regime, supporting the arguments given in \cite{hubeny}.
The QNM for toroidal, cylindrical or planar black holes (in anti-de
Sitter space) are quite similar to those of the Schwarzschild-anti-de
Sitter black hole \cite{hubeny,cardoso2}.

\vskip 3mm

\vskip .5cm

\section*{Acknowledgments} This work was partially funded
by Funda\c c\~ao para a  Ci\^encia e Tecnologia (FCT) 
through project SAPIENS 36280. VC  also 
acknowledges finantial support from FCT 
through PRAXIS XXI programme.
JPSL thanks Observat\'orio Nacional do Rio de Janeiro for
hospitality. 


\end{document}